\documentclass[prd,twocolumn,superscriptaddress,longbibliography,preprintnumbers,floatfix,nofootinbib]{revtex4-1}

\usepackage{url}
\usepackage{xspace}
\usepackage{dsfont}
\usepackage{amssymb}
\usepackage{amsmath}
\usepackage{graphicx}
\usepackage[caption=false]{subfig}
\usepackage[colorlinks=true,citecolor=blue]{hyperref}
\usepackage{multirow}
\usepackage{hhline}
\usepackage{xcolor}
\usepackage{comment}

\begin{document} 

\title{Generator Based Inference (GBI)}

\author{Chi Lung Cheng}
\email{ccheng84@wisc.edu}
\affiliation{Department of Physics, University of Wisconsin, Madison, WI 53706, USA}
\affiliation{Physics Division, Lawrence Berkeley National Laboratory, Berkeley, CA 94720, USA}

\author{Ranit Das}
\email{ranit@physics.rutgers.edu}
\affiliation{New High Energy Theory Center, Rutgers University Piscataway, NJ 08854, USA}

\author{Runze Li}
\email{runze.li@yale.edu}
\affiliation{Department of Physics, Yale University,  New Haven, CT 06511, USA}

\author{Radha Mastandrea}
\email{rmastand@berkeley.edu}
\affiliation{Department of Physics, University of California, Berkeley, CA 94720, USA}
\affiliation{Physics Division, Lawrence Berkeley National Laboratory, Berkeley, CA 94720, USA}

\author{Vinicius Mikuni}
\email{vmikuni@lbl.gov}
\affiliation{National Energy Research Scientific Computing Center, Berkeley Lab, Berkeley, CA 94720, USA}

\author{Benjamin Nachman}
\email{bpnachman@lbl.gov}
\affiliation{Physics Division, Lawrence Berkeley National Laboratory, Berkeley, CA 94720, USA}
\affiliation{Department of Particle Physics and Astrophysics, Stanford University, Stanford, CA 94305, USA}
\affiliation{Fundamental Physics Directorate, SLAC National Accelerator Laboratory, Menlo Park, CA 94025, USA
}

\author{David Shih}
\email{dshih@physics.rutgers.edu}
\affiliation{New High Energy Theory Center, Rutgers University Piscataway, NJ 08854, USA}

\author{Gup Singh}
\email{gupsingh@berkeley.edu}
\affiliation{Physics Division, Lawrence Berkeley National Laboratory, Berkeley, CA 94720, USA}

\begin{abstract}
Statistical inference in physics is often based on samples from a generator (sometimes referred to as a ``forward model") that emulate experimental data and depend on parameters of the underlying theory.  Modern machine learning has supercharged this workflow to enable high-dimensional and unbinned analyses to utilize much more information than ever before.  We propose a general framework for describing the integration of machine learning with generators called Generator Based Inference (GBI).  A well-studied special case of this setup is Simulation Based Inference (SBI) where the generator is a physics-based simulator.  In this work, we examine other methods within the GBI toolkit that use data-driven methods to build the generator.  In particular, we focus on resonant anomaly detection, where the generator describing the background is learned from sidebands.  We show how to perform machine learning-based parameter estimation in this context with data-derived generators.  This transforms the statistical outputs of anomaly detection to be directly interpretable and the performance on the LHCO community benchmark dataset establishes a new state-of-the-art for anomaly detection sensitivity.

\end{abstract}

\maketitle
\flushbottom

\section{Introduction}
\label{sec:intro}

Simulations are a critical component of inference in particle physics and beyond.  The typical protocol for estimating parameters from data is to simulate data with a number of parameter values, identify one (or at most a few) key observables, construct histograms of those observables, and then find the one that best matches data using a $\chi^2$ or similar metric.  This approach is fundamentally limiting because it requires a significant and lossy reduction in the dimension of the data to compare data with simulation.  Recent advances in machine learning have enabled a new paradigm of unbinned and high-dimensional inference using simulation-based inference (SBI)~\cite{Cranmer_2020}.

There are a number of strategies for approaching parameter estimation with SBI.\footnote{We will use SBI and \textit{machine learning-based SBI for parameter estimation} synonymously, even though SBI does not always involve machine learning and there are other such tasks aside from parameter estimation, including unfolding~\cite{Arratia:2021otl,Huetsch:2024quz}.}  For example, one approach is to re-purpose machine learning classifiers as likelihood ratio estimators for constructing frequentist confidence intervals.  This strategy was recently deployed on real data to increase the precision on the interference effects in Higgs boson decays to four leptons~\cite{ATLAS:2024ynn,ATLAS:2024jry}.  A central aspect of these strategies is that they require a set of simulated signal and background events to train the classifier.  This is only possible in a limited number of final states where the background prediction comes directly from physics simulations.  In most analyses, some component of the background is estimated directly from data.  Also, most analyses target one (and at most a few) signal process(s).  These signal model(s) serve as benchmarks for optimizing signal-sensitive region(s).

We propose a generalized framework that enables unbinned and high-dimensional inference to extend beyond the SBI setting.  We call this new toolkit Generator Based Inference (GBI).  While the generator used for inference could come from simulation (SBI is a type of GBI), it may also be data-driven or a data-simulation hybrid.  Nearly any data-driven method for estimating background probability densities can be promoted to an unbinned and high-dimensional likelihood using machine learning. Recently, a number of machine learning-based methods for background estimation have been proposed in the context of resonant anomaly detection (see~\cite{Kasieczka:2021xcg,Aarrestad:2021oeb,Karagiorgi:2021ngt,Golling:2023yjq,Belis:2023mqs} for reviews and original references), where sideband methods can be used to approximate the background. Here we will illustrate the idea of GBI by applying it to resonant anomaly detection, by repurposing these data-driven background templates to infer the parameters of the underlying signal model.

Previous resonant anomaly detection methods trained classifiers or directly estimated likelihood ratios to distinguish the surrogate background from data in the signal mass window.  Without a particular signal hypothesis, these analyses performed two-sample tests to compute $p$-values.  These tests were mostly based on parametric fits to the mass distribution after event selections. By contrast, in GBI, we will directly use estimates for the background and signal likelihoods to infer the signal strength and other signal parameters.

For learning the signal likelihood, we first consider the Residual Anomaly Detection with Density Estimation (R-ANODE) method~\cite{Das:2023bcj}.  In this approach, the signal is completely learned from the data and the extraction of signal strength is performed by marginalizing over thousands of neural network parameters.
We demonstrate that the R-ANODE method is also capable of constructing confidence intervals for the signal strength from the results and we explore sources of bias.  In practice and with current tools, the flexible signal form for R-ANODE absorbs any mis-modelings/statistical fluctuations of the background, which sets a floor on the signal sensitivity.

To address the challenge of signal simulation flexibility, we consider a second, data-simulation hybrid approach.  The signal likelihood (ratio) is parameterized from a set of physics models specified by parameters $\theta$.  The neural network parameters specifying the likelihood ratio are pre-trained using signal simulation combined with the data-driven background model and then the fit to data only allows for the physical parameters to vary.  Since the background mis-modeling is mostly not signal-like, this extension of the Prior-Assisted Weak Supervision (PAWS) method is largely insensitive to background simulation defects~\cite{Cheng:2024yig}.  We show that changing the loss function for PAWS from a standard weakly-supervised approach to a GBI approach does not reduce the sensitivity, despite the added interpretatiblity of the result in terms of confidence intervals on the signal strength and the physics model parameters.  Along the way, we also significantly enhance the sensitivity to PAWS, showing the ability to detect anomalies  starting with an initial signal injection of 0.1$\sigma$.

This paper is organized as follows.  Section~\ref{sec:methods} introduces the statistical and machine learning aspects of GBI in the context of resonant anomaly detection.  The LHC Olympics dataset used for benchmarking is briefly described in Sec.~\ref{sec:data}.  Numerical results are presented in Sec.~\ref{sec:results} and the paper ends with conclusions and outlook in Sec.~\ref{sec:conclusions}.

\section{Methods}
\label{sec:methods}

In resonant AD, each event is specified by a mass $m$ and other features $x$ that will be used to find the signal. A particular interval in $m$ defines a signal region (SR) and the complement of the SR is called the sideband region (SB). We can model the data in the SR as a mixture model: $p_D(x)=\mu p_S(x)+(1-\mu)p_B(x)$, where $p_S$ and $p_B$ are the probability densities of the signal and background processes, respectively. The primary goal in the search for new physics is to perform parameter estimation on the signal fraction $\mu$. When $\mu$ is not consistent with zero, we have evidence for a discovery. If $\mu$ is non-zero, we subsequently would like to learn about the properties of the anomalous events. 

The first step is to estimate $p_B$ using the SB data, which is assumed to have negligible amount of signal. Following the Anomaly Detection with Density Estimation (ANODE)~\cite{Nachman:2020lpy} and Classifying Anomalies THrough Outer Density Estimation (CATHODE)~\cite{Hallin:2021wme} methods, we approximate $p_B(x|m)$ from the SB using conditional generative models.  One of our methods (R-ANODE) requires a direct density estimation and so we use a conditional normalizing flow (NF)~\cite{rezende2016variationalinferencenormalizingflows} while our other method (PAWS) does not require density estimation and so we use Conditional Flow-Matching (CFM)~\cite{lipman:2022flow} to achieve better sample quality. NFs are a class of deep neural networks that map a simple random variable to the data distribution and, as functions, integrate to unity, allowing for both probability density estimation and sampling. CFMs are an efficient way to train a Continuous Normalizing Flow (CNF) by learning the vector field that generates a continuous transformation between the random variable and data. We train a single CFM model for PAWS and find that its precision is not limiting for the final sensitivity. For R-ANODE we combine multiple NF models to improve accuracy.  


The next step is to specify the signal simulation used to construct $p_S$.

\subsection{Non-parametric Signal}

As a first approach, we make minimal assumptions about the functional form of $p_S$. Following the Residual ANODE (R-ANODE) approach~\cite{Das:2023bcj}, we fit a normalizing flow $f(x)$ in the SR by freezing the background density $p_B(x)$ learned in the SB and maximizing the following likelihood over $\mu$ and the parameters of $f$:
\begin{align}
\label{eq:ranode}
    \sum_{x_i\sim p_D} \log(\mu f(x_i)+(1-\mu) p_B(x_i)) \,,
\end{align}
where\footnote{Both random variables and their realization are depicted with the same lower case letters as the meaning is clear from context.} $f(x)$ represents the signal density $p_S(x)$. To improve performance, $p_B$ comes from the ensemble average of 20 NF models trained in the SB with background events only.\footnote{In practice, the SB would have a small amount of signal from the finite width of the resonance.  We found no impact from this signal leakage and it significantly complicates the workflow because the NFs need to be retrained for every signal model and for every amount of signal injection, so it is henceforth dropped.} For $f(x)$, at any given amount of signal injection, we scan over 20 $\mu$ values ranging from $10^{-5}$ to $10^{-1}$ in log scale. For each $\mu$, we first train an ensemble of 20 NF models $f(x)$ with varying model initializations and train/validation dataset splittings, while keeping the same test dataset. The per-event likelihood is averaged over the ensemble models $f(x)$ on the test dataset.  Then we pick the $\hat{\mu}$ that maximizes the likelihood of Eq.~\ref{eq:ranode} on the test dataset, using a Gaussian process regressor to interpolate between scan points, with uncertainty approximated by where the log likelihood decreases by $\chi^2_{\alpha,1}$, the value from a $\chi^2$ table with confidence level $\alpha$ and one degree of freedom~\cite{Wilks:1938dza}. 

It is too computationally expensive to validate these uncertainties numerically with R-ANODE, as we do in the next section for PAWS. 

\subsection{Incorporating Physical Priors}

While minimal assumptions yield broad sensitivity, stronger sensitivity can be achieved by restricting the functional space used to model the signal. This is the central idea of PAWS~\cite{Cheng:2024yig}, where a class of signal models with parameters $\theta$ are used to pre-train a set of neural network-based likelihood ratios. Then, the resulting parameterized likelihood is fit to data with the neural network weights and biases frozen, allowing only the physical parameters $\theta$ and the signal fraction $\mu$ to vary.

The pre-trained model is a fully supervised parameterized classifier~\cite{Cranmer:2015bka,Baldi:2016fzo} $g(x,\theta)$ trained with the cross-entropy loss:

\begin{align}
\label{eq:bce}
    -\sum_{\substack{\theta_i\sim p(\theta)\\ x_i\sim p_S(x|\theta_i)}} \log(g(x_i,\theta_i)) - \sum_{\substack{\theta_i\sim p(\theta)\\x_i\sim p_B}} \log(1 - g(x_i,\theta_i))\,.
\end{align}
The parameter values $\theta_i$ assigned in the second sum follow the same prior $p(\theta)$ but do not affect $p_B(x)$ (learned from data as described at the beginning of this section), ensuring $\theta$ itself is not a useful feature for classification. To enhance the accuracy and stability of the likelihood ratio estimate, we train an ensemble of 10 such prior models, each using a different data realization and model initialization. The average output $\langle g(x,\theta) \rangle$ across the ensembles is used to estimate the optimal classifier $g^*(x,\theta)$. The learned $g^*$ approximates the likelihood ratio:

\begin{align}
    g^*&\approx\kappa(\theta)\,\frac{p_\text{S}(x|\theta)}{p_\text{S}(x|\theta)+p_\text{B}(x)}\,,
\end{align}
where $\kappa(\theta)\equiv N_B/N_S(\theta)$ accounts for the relative number of signal and background events in the training. We fix the $\theta$ dependence by generating more events per signal mass point than are needed and then dropping events to counter-balance acceptance effects that result in slightly uneven numbers of signal events as a function of the particle masses (new with respect to Ref.~\cite{Cheng:2024yig}). The fully supervised likelihood ratio is extracted as $\Lambda_\text{FS}(x|\theta) \equiv p_\text{S}(x|\theta)/p_\text{B}(x)\approx \kappa(\theta)\,\langle g^*(x,\theta) \rangle/(1-\langle g^*(x,\theta) \rangle)$. From this, we can build the weakly-supervised likelihood ratio between data and the background-only hypothesis:
\begin{align}\nonumber
\Lambda_\text{WS}(x|\theta,\mu)\equiv\frac{p_D(x|\theta,\mu)}{p_B(x)}&=\frac{\mu p_S(x|\theta)+(1-\mu)p_B(x)}{p_B(x)}\\
&=\mu\Lambda_\text{FS}(x|\theta)+(1-\mu)\,. \label{eq:lambda_ws}
\end{align}

Lastly, we maximize the likelihood (ratio) over $\theta$ and $\mu$:
\begin{align}
\label{eq:bceWS}
    \sum_{x_i\sim p_D} \log(\Lambda_\text{WS}(x_i|\theta,\mu)) \,.
\end{align}
Note that unlike standard weakly supervised learning (including the original PAWS formulation), Eq.~\ref{eq:bceWS} only requires one dataset and not two - there is no direct comparison between data and a reference background sample.  Everything that we need to know about the background is encoded into $\Lambda_\text{WS}$. 
Furthermore, Eq.~\ref{eq:bceWS} is highly constrained and we find no evidence for overfitting.  We thus extend the previous PAWS results by training and testing with the same data. This increases the sensitivity by using more data for the fit.  As with the R-ANODE case, we estimate uncertainties by where the log likelihood decreases by $\chi^2_{\alpha,k}$ for confidence level $\alpha$ and $k$ fitted parameters~\cite{Wilks:1938dza}.  Since we want to study $\theta$ in addition to $\mu$ (when it is non-zero), it is useful to derive confidence intervals across all parameters.  Building confidence regions is challenging in more than one dimension, but we can approximate it with the inverse square root of the Fisher information~\cite{wald1943}. Since there are so few parameters to fit, it is also possible to check the coverage using pseudoexperiments.  We report uncertainties using all three methods.

\section{Dataset}
\label{sec:data}

We use an extended version of the LHC Olympics dataset~\cite{Kasieczka:2021xcg,LHCOlympics} that is similar to the one created for Ref.~\cite{Cheng:2024yig}. In particular, the background consists of generic quark-gluon scattering processes to produce dijet events and the signal is $W' \rightarrow X Y $ with $X, Y \rightarrow q\bar{q}$.  The parent mass is 3.5\,TeV and we generate a grid of $\theta=(m_X,m_Y)$ values with $m_X,m_Y<600$\,GeV in increments of 50\,GeV and consisting of 100,000 events per sample. The upper mass bound ensures that the quarks are well-contained within a single jet.  All processes were simulated using \textsc{Pythia}~8.219~\cite{Sjostrand:2006za, Sjostrand:2014zea} and \textsc{Delphes}~3.4.1~\cite{deFavereau:2013fsa, Mertens:2015kba}.

The reconstructed particles of each event are clustered into jets using the anti-$k_{T}$ algorithm~\cite{Cacciari:2005hq, Cacciari:2011ma, Cacciari:2008gp} with $R=1.0$.  All events are required to satisfy $p_{T} >$ 1.2\,TeV.  The two highest $p_T$ jets $J_1$ and $J_2$ are sorted such that $m_{J_1} > m_{J_2}$.  Each event is characterized by the same six features ($M=6$) that have been widely used in LHCO studies: $x=\{m_{J_1}, m_{J_2}, \tau_{21}^{J_1}, \tau_{21}^{J_2}, \tau_{32}^{J_1}, \tau_{32}^{J_2}\}$, where the $n$-subjettiness ratios are defined as $\tau_{ij} \equiv \tau_{i} /\tau_{j}$~\cite{Thaler:2010tr,Thaler:2011gf}.  These are sensitive to the two- ($\tau_{21}$) or three- ($\tau_{32}$) prong nature of the jets.  While PAWS uses all features, R-ANODE uses $\{m_{J_2},m_{J_1}-m_{J_2},\tau_{21}^{J_1}, \tau_{21}^{J_2}\}$.

The signal is resonant in $m_{JJ}$ and a full analysis would scan over $m_{JJ}$ to form signal regions and sideband regions.  As in many previous studies, for simplicity, we focus on the one signal region $m_{JJ} \in [3.3, 3.7]$\,TeV that contains the signal.  The sideband $m \not\in [3.3, 3.7]$\,TeV is used to estimate the background-only distribution of features $p_\text{B}(x|m)$.  We additionally estimate $p(m)$ using a parametric fit to the $m_{JJ}$ distribution.  This is used for the $p_B(x)$ estimation in R-ANODE and to sample background examples in PAWS in the SR. For the latter, we sample 3 times more background events as are expected in data and use those for the pre-trained model.


For illustration, we focus on the LHC Olympics benchmark signal $\theta=(100, 500)$ GeV and show the behavior across mass values when relevant.   Ref.~\cite{Cheng:2024yig} showed that PAWS remains sensitive to signal models that are not exactly in the pre-training set, but it would be interesting to explore the depth-breadth tradeoff between R-ANODE and PAWS in future work with different signal models and likely different feature sets.

For robustness, all results are presented as medians over 20 data ensembles, generated by randomly selecting signal events for a given injection.

\section{Results}
\label{sec:results}

The goal of GBI is parameter estimation and so we are focused on the discovery inference (determining $\mu$) and the interpretability followup (determining the other parameters).  Many previous AD papers focused on classification performance as an intermediate step of the full analysis pipeline.  At the level of classification performance, we did not observe any degradation from the GBI version over the original PAWS implementation.  The only difference between these two approaches is the loss function, where the GBI version has only one dataset and uses Eq.~\ref{eq:bceWS} while the original version used two datasets and used cross entropy between data and the background reference samples. With respect to the original PAWS paper, we also extend the sensitivity on $\mu$ by a factor of approximately five from removing the data splitting and ensembling the prior models. As for binned template fits, we did not find evidence for overfitting and the increased statistics for training and testing improved the performance.

\begin{figure}[ht!]
    \centering
    \includegraphics[width=0.99\linewidth]{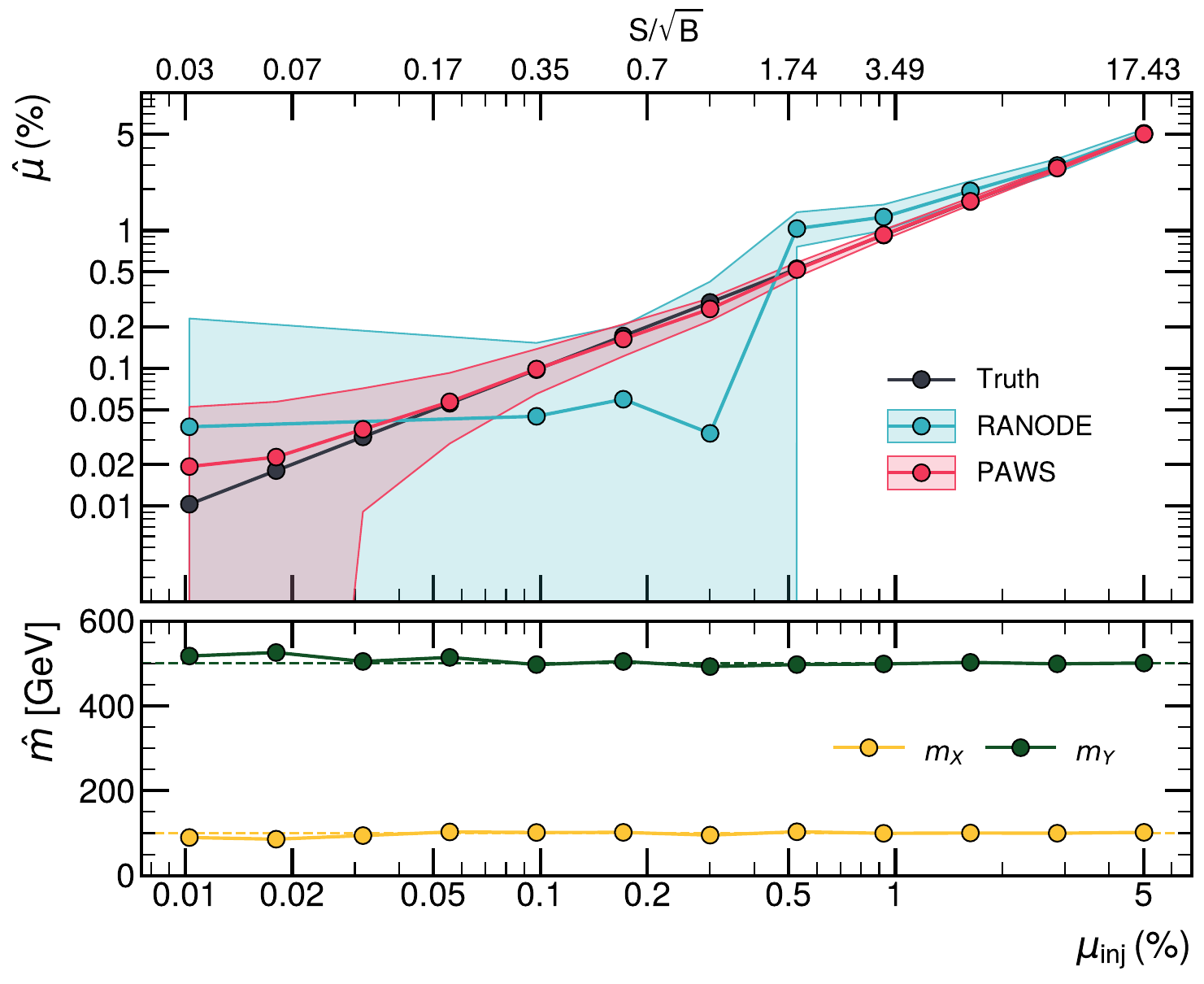}
    \caption{Median estimated signal fraction $\hat{\mu}$ using R-ANODE and GBI-PAWS methods, shown as a function of injected signal fraction $\mu_{\text{inj}}$ for the $(m_X, m_Y) = (100, 500)$\,GeV benchmark signal. The black line denotes the truth ($\hat{\mu} = \mu_{\text{inj}}$). The shaded bands are median $2\sigma$ confidence intervals from profile likelihood scans.  For simplicity, only the likelihood scan uncertainties are shown for GBI-PAWS, but all methods agree as shown in Fig.~\ref{fig:massscan}.  The maximum likelihood estimate $\hat{m}$ (from the ensemble closest to the median $\hat{\mu}$) for GBI-PAWS is shown in the lower panel.}
    \label{fig:mu_inference}
\end{figure}

The discovery inference is presented in Fig.~\ref{fig:mu_inference}.  We present the inferred $\mu$, with its corresponding confidence intervals, for both R-ANODE and GBI-PAWS.
A signal is discovered when the uncertainty band is not consistent with zero.  We studied many variations on the setup in order to arrive at the final configuration presented here.  For R-ANODE, we found that significantly reducing the number of neural network parameters with respect to Ref.~\cite{Das:2023bcj} improved the bias at low signal injection.  Appendix~\ref{sec:ranodebias} shows that by training the background flow directly on signal region background (not possible in practice) and by sampling from a trained flow and pretending it is data (so the probability density is known exactly), we found that most of the artifacts degrading performance at low signal injection come from the interpolation of the background model.  In the case of GBI-PAWS, we found the model sometimes learned a wrong mass value at low signal injection due to a background peak around $\{m_X, m_Y\} = \{220,\,75\}$\,GeV mimicking a low-mass signal. To steer the fit away from this background artifact, an exponential penalty term is added to the loss function when the fitted mass parameters fall below a threshold of 85\,GeV. It may also be possible to mitigate this in the future by using additional distinguishing features in the training so that the signal and background densities are less likely to overlap ($\tau_{12}$ and $\tau_{23}$ are not enough in this case). With these configurations, both methods are able to track the true $\mu$ with the fitted $\mu$ within uncertainty.

Focusing on the LHCO benchmark signal, the R-ANODE approach is able to find anomalies that start above about $1\sigma$ and the GBI-PAWS approach is able to find signals that start above about $0.1\sigma$.  These two approaches are complementary - the GBI-PAWS approach can find rarer signals than R-ANODE, but requires more assumptions.  Even though the R-ANODE inference includes thousands of parameters, the precision and accuracy marginalized over the neural network are competitive with respect to the inclusive search.

\begin{figure}[ht!]
    \centering
    \includegraphics[width=0.95\linewidth]{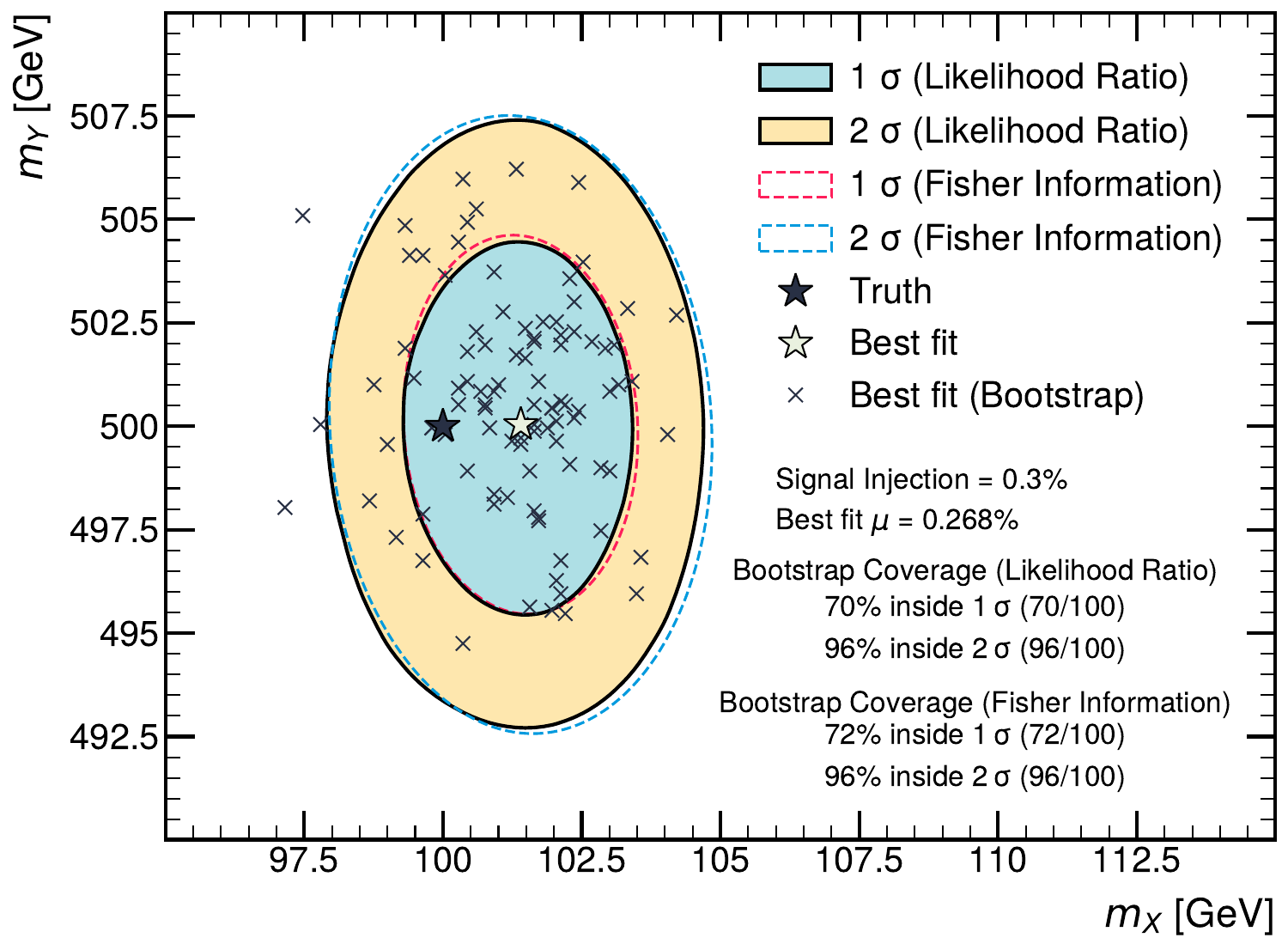}
    \includegraphics[width=0.95\linewidth]{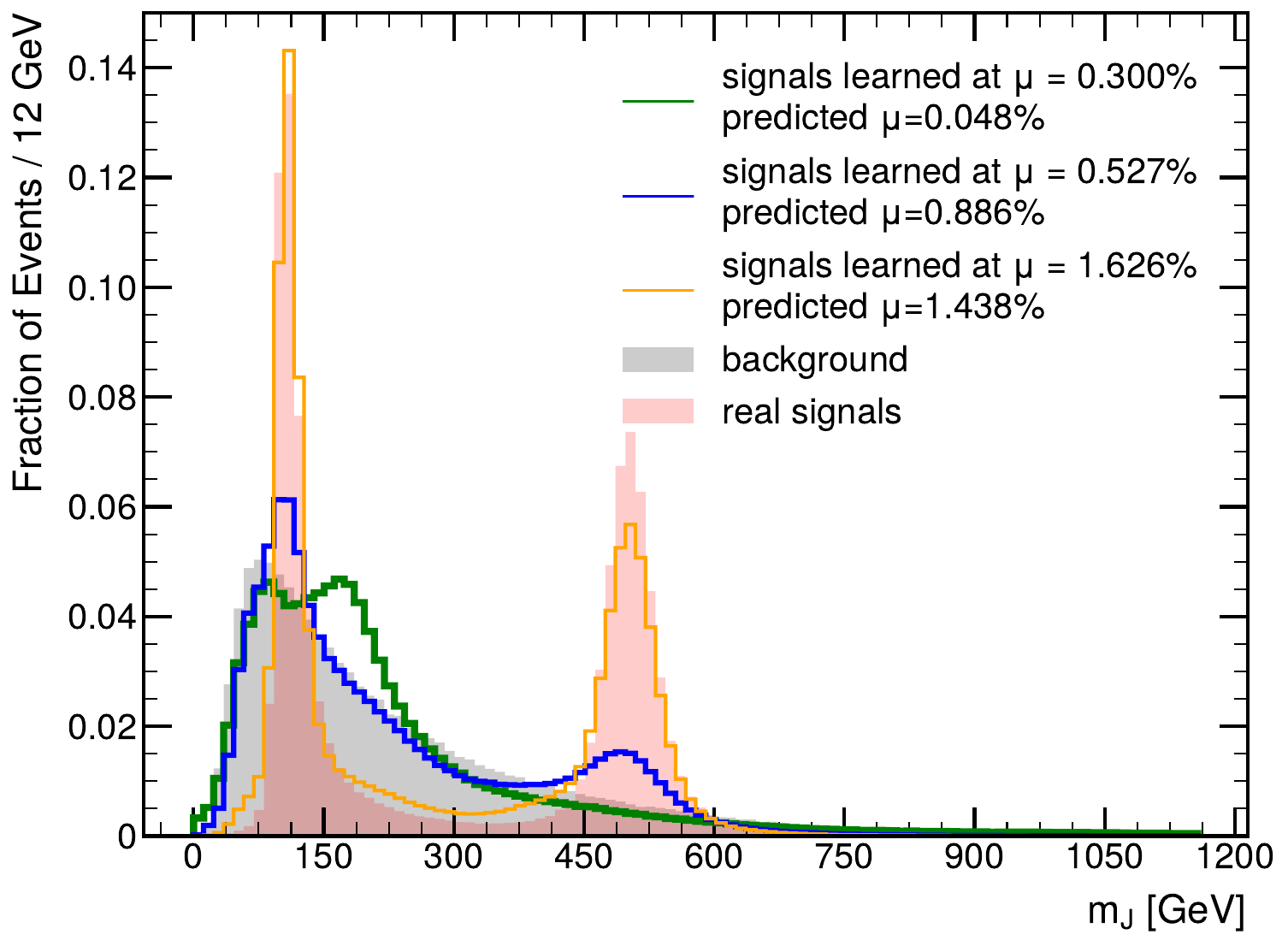}
    \caption{Top: Inferred anomaly mass properties ($m_X$, $m_Y$) from GBI-PAWS. Bottom: Jet mass distribution from R-ANODE, illustrating how the learned signal density evolves with increasing injected signal strength, alongside the true signal (gray).}
    \label{fig:mass_contours}
\end{figure}

For a given discovery, we can examine the non-$\mu$ parameters to interpret what has been found.  In the case of GBI-PAWS, this would be the fitted mass values and in the case of R-ANODE, this would be the learned signal $p_S$.  Figure~\ref{fig:mass_contours} presents these quantities for the LHCO benchmark.  GBI-PAWS is able to identify the correct mass values and we validate the coverage of each uncertainty quantification method (likelihood ratio and Fisher information) is correct through pseudoexperiments.  In the R-ANODE case, when $\mu$ is significantly away from zero, $p_S$ has peaks at the correct mass values.

\begin{figure}[ht!]
    \centering
    \includegraphics[width=0.95\linewidth]{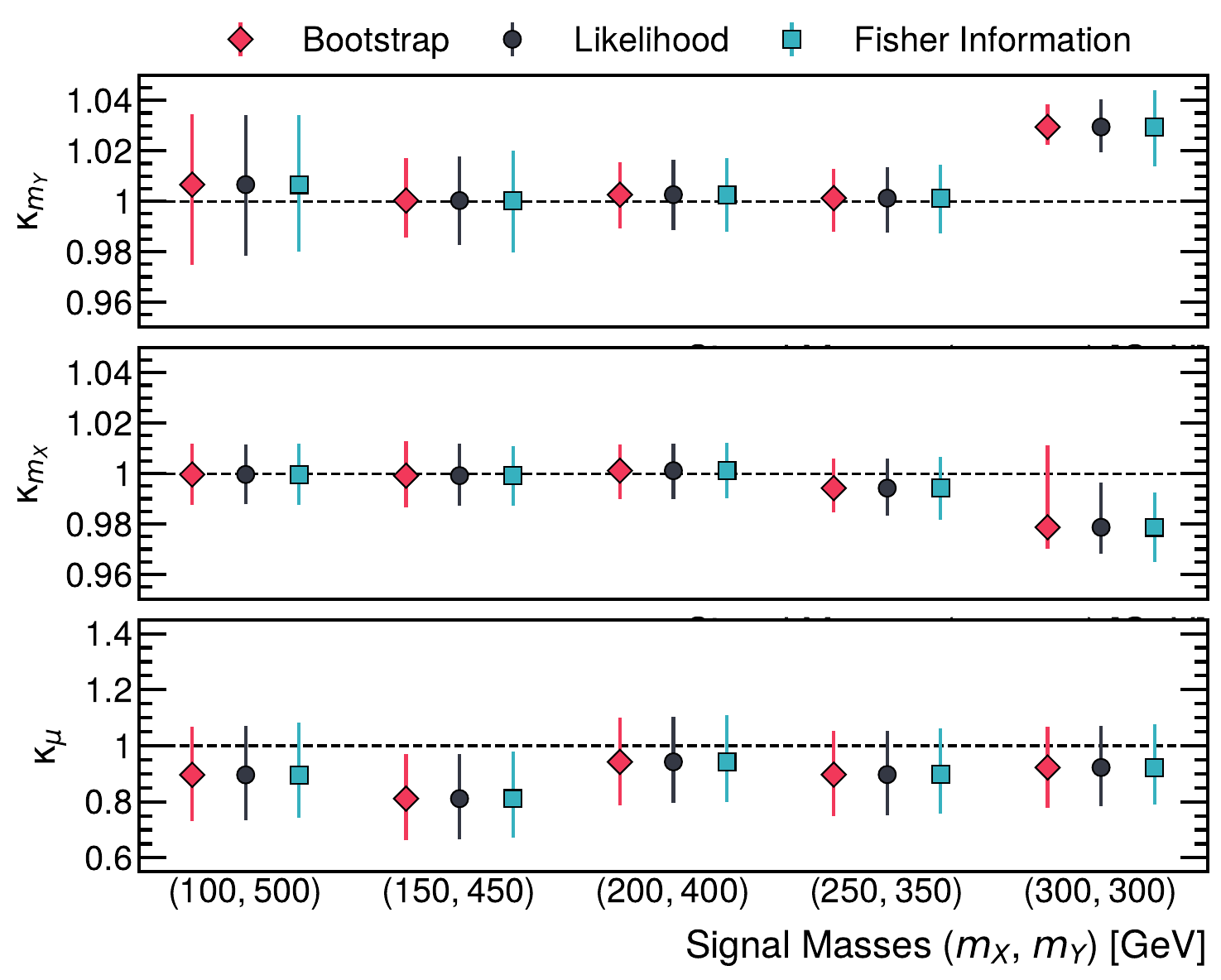}
    \caption{Fitted parameter values ($\mu$, $m_X$, $m_Y$) divided by true values (denoted $\kappa$) for GBI-PAWS, evaluated across multiple signal models. Uncertainties are estimated via bootstrap (from 1000 pseudoexperiments), likelihood ratio profiling, and Fisher information matrix approaches. The true signal fraction is fixed at 0.3\%.}
    \label{fig:massscan}
\end{figure}

Lastly, we consider the performance across signal models.  R-ANODE requires no assumption on the signal and we find that its performance is similar across masses. Scanning signal models extensively for R-ANODE is also computationally prohibitive.  In contrast, GBI-PAWS requires significant prior information, but is much faster. Fig.~\ref{fig:massscan} shows that the GBI-PAWS works across mass values.  In particular, we find that the method is effective across the entire mass grid and the three uncertainty quantification methods agree.  We observe a slight bias only for the (300,300) GeV mass point, but the shift is a few percent and is likely practically irrelevant for the interpretability task.

\section{Conclusions}
\label{sec:conclusions}

In this paper, we have introduced Generator Based Inference (GBI) -- a new strategy to combine machine learning with data that generalizes Simulation Based Inference.  Given the importance of data-driven background estimates, this framework  is essential for widespread adaptation of likelihood-free methods, enabling unbinned and high-dimensional analyses that enhance discovery potential.  We have demonstrated this approach using resonant anomaly detection, where the background can be estimated from sideband information.  Two GBI approaches are studied: R-ANODE, which learns both the signal and background generators from the data and GBI-PAWS, which learns the parameters of the signal from simulation and the full background generator from the data.  R-ANODE was proposed in Ref.~\cite{Das:2023bcj} and we extend it by constructing confidence intervals from the results and by exploring sources of bias. PAWS was proposed in Ref.~\cite{Cheng:2024yig} and we have introduced a number of innovations: the background is estimated directly from data, the loss function only requires the data (and not also a reference sample), confidence intervals are extracted from the result, and the overall sensitivity is significantly improved without data splitting.  These two methods are complementary in terms of their breadth and depth and serve as templates for future GBI approaches.  

\section*{Data and code availability}

The original LHC Olympics datasets are available on Zenodo.  Our additional signal models, with the corresponding scripts to generate them, are also on Zenodo. The code we used to train and evaluate our models is available at \url{https://github.com/hep-lbdl/paws-sbi}.

\vspace{1cm}

\section*{Acknowledgments}

C.L.C., R.M., V.M, and B.N. are supported by the U.S.~DOE Office of Science under contract DE-AC02-05CH11231.  R.D. and D.S. are supported
by DOE grant DE-SC0010008. R.L. is supported by DOE grant DE-SC0017660. R.M. and B.N. are additionally supported by the John Templeton Foundation.  This research used resources of the National Energy Research Scientific Computing Center, a DOE Office of Science User Facility supported by the Office of Science of the U.S. Department of Energy under Contract No. DE-AC02-05CH11231 using NERSC award HEP-ERCAP0021099.

\bibliography{main}
\bibliographystyle{JHEP}

\clearpage

\appendix

\section{Model Bias in R-ANODE}
\label{sec:ranodebias}
As shown in Fig.~\ref{fig:mu_inference}, at low signal strength $(S/\sqrt{B} < 4)$, the estimated signal fraction $\hat{\mu}$ tends to bias towards a larger value, due to a combined effect from both likelihood fitting uncertainty and imperfection in background model $p_B$.

\begin{figure}[ht!]
    \centering
    \includegraphics[width=0.95\linewidth]{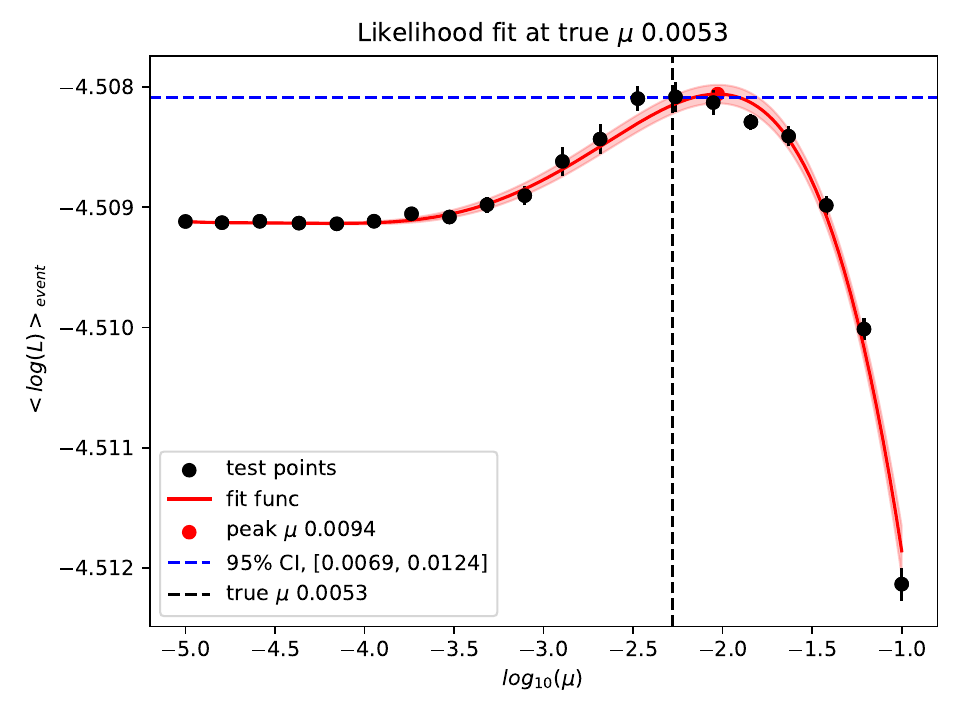}
    \includegraphics[width=0.95\linewidth]{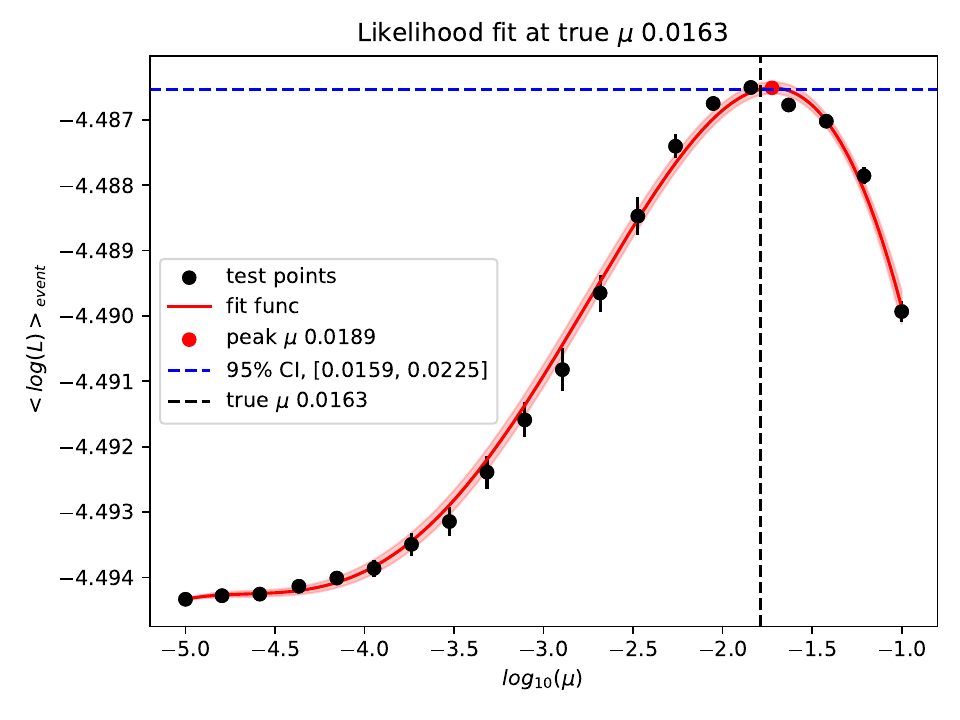}
    \caption{R-ANODE likelihood fitting at two different true $\mu$ values.}
    \label{fig:ranode_fitting}
\end{figure}

For R-ANODE, as described in section \ref{sec:methods}, we use an ensemble of 20 $f(x)$ and 20 $p_B$ models to get the average likelihood per event. The average values of all event's likelihood at each of the 20 $\mu$ test points are calculated, and these 20 ($\mu$, $<L(\mu)>$) pairs are fitted with a gaussian process regressor in order to find the $\hat{\mu}$ that maximize the likelihood. In order to stabilize the fitting, we also perform a bootstrapping of $f(x)$ models to estimate the uncertainty that goes into the fitting. Fig.~\ref{fig:ranode_fitting} illustrates fitting at two distinct injected signal strengths. We see that at higher true $\mu$, the peak of fitted curve is closer to the correct value compared to the case with low signal strength. This could be caused by both the flat tail distribution at low test $\mu$ and the uncertainty in $<L(\mu)>$. These effects can be compensated by increasing the size of model ensemble, adding more test points in scan, or doing a second round of scan focusing explicitly around the peak. Due to the computational cost, these effects are not thoroughly explored in this study.

\begin{figure}[ht!]
    \centering
    \includegraphics[width=0.95\linewidth]{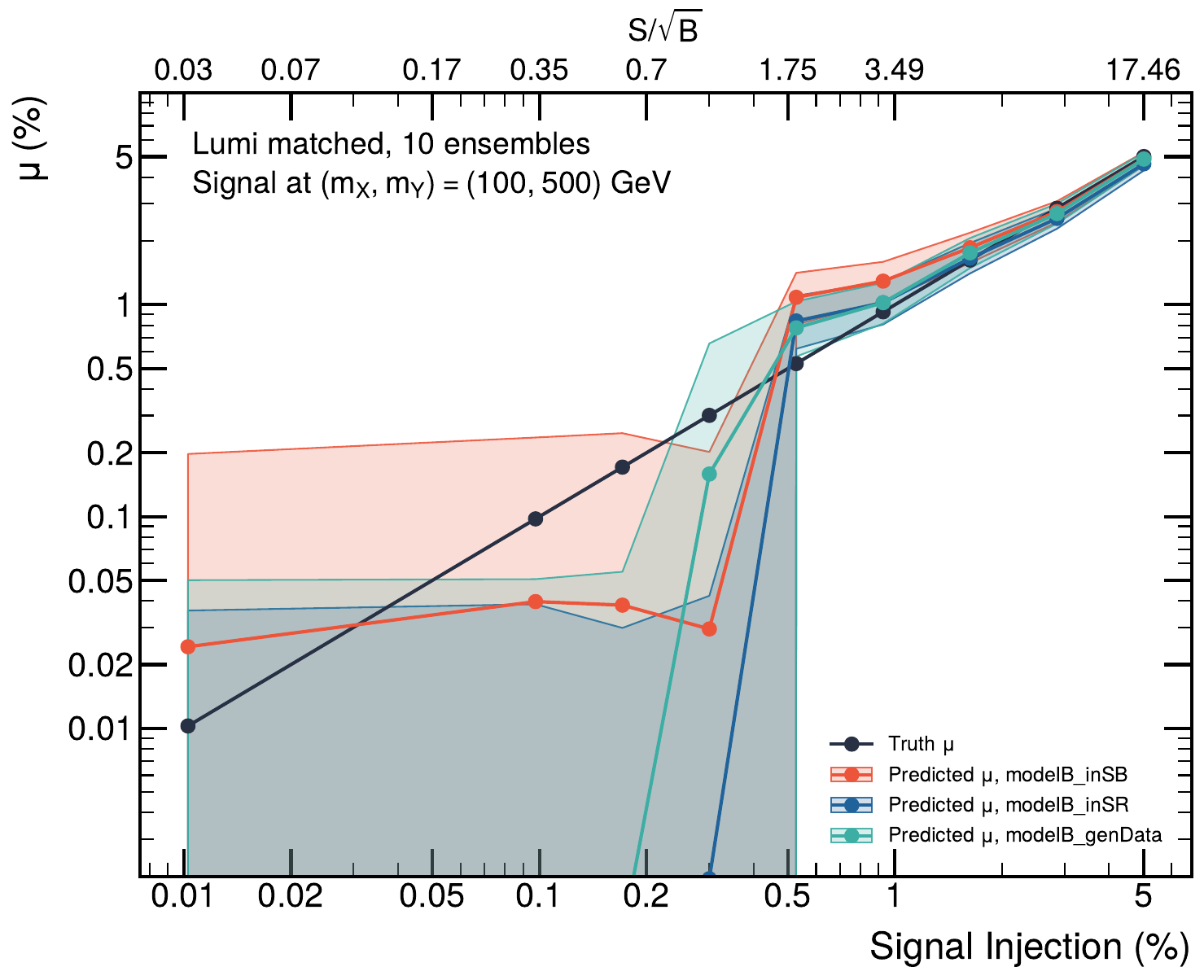}
    \caption{Median estimated signal fraction $\hat{\mu}$ using R-ANODE where background model $p_B$ comes from: 1. SB interpolation 2. Pure background events in SR 3. Directly use $p_B$ to generate backgrounds in SR data.}
    \label{fig:ranode_comparison_modelB}
\end{figure}

Another source of bias comes from the imperfection of background model $p_B$. In our study, $p_B$ is an ensemble of 20 normalizing flow models trained in SB and interpolated into SR. Due to the interpolation, the distribution carried by the model $p_B$ is different from the true underlying distribution of backgrounds in SR. As a result, what model $f(x)$ learn becomes a mixture of signal samples and the interpolation error from model $p_B$. When signal injection is large, model $f(x)$ will more likely learn the signal distribution, but when signal injection is small, model $f(x)$ will learn the error of model $p_B$, and this error will be interpreted as signals. This effect causes the inferred $\hat{\mu}$ to be larger than true $\mu$. In order to address this effect, we try to minimize the interpolation error of model $p_B$ by training $p_B$ directly using background events in SR (not possible in practice), or replacing backgrounds in our data with samples generated by $p_B$ (so the probability is known exactly). The results are compared in Fig.~\ref{fig:ranode_comparison_modelB}. We find that both training model $p_B$ with background events in SR and using $p_B$ to sample background events in data reduce the bias of $\hat{\mu}$. Especially, when the signal strength is lower than the discovery limit, both of the two method returns $\hat{\mu}$ equals 0. The remaining bias here could be caused by the fitting effect.

It is also worth noting that the size of model $f(x)$ has a large impact on $\hat{\mu}$ bias, since model $f(x)$ has more parameters can fit on the error of model $p_B$ easier. In our study, we use model $f(x)$ with $\frac{1}{10}$ number of parameters than used in Ref.~\cite{Das:2023bcj}, and in real analysis with R-ANODE hyperparameter tuning in control region might be necessary.

\section{R-ANODE Performance Across Signal Models}

\begin{figure}[ht!]
    \centering
    \includegraphics[width=0.95\linewidth]{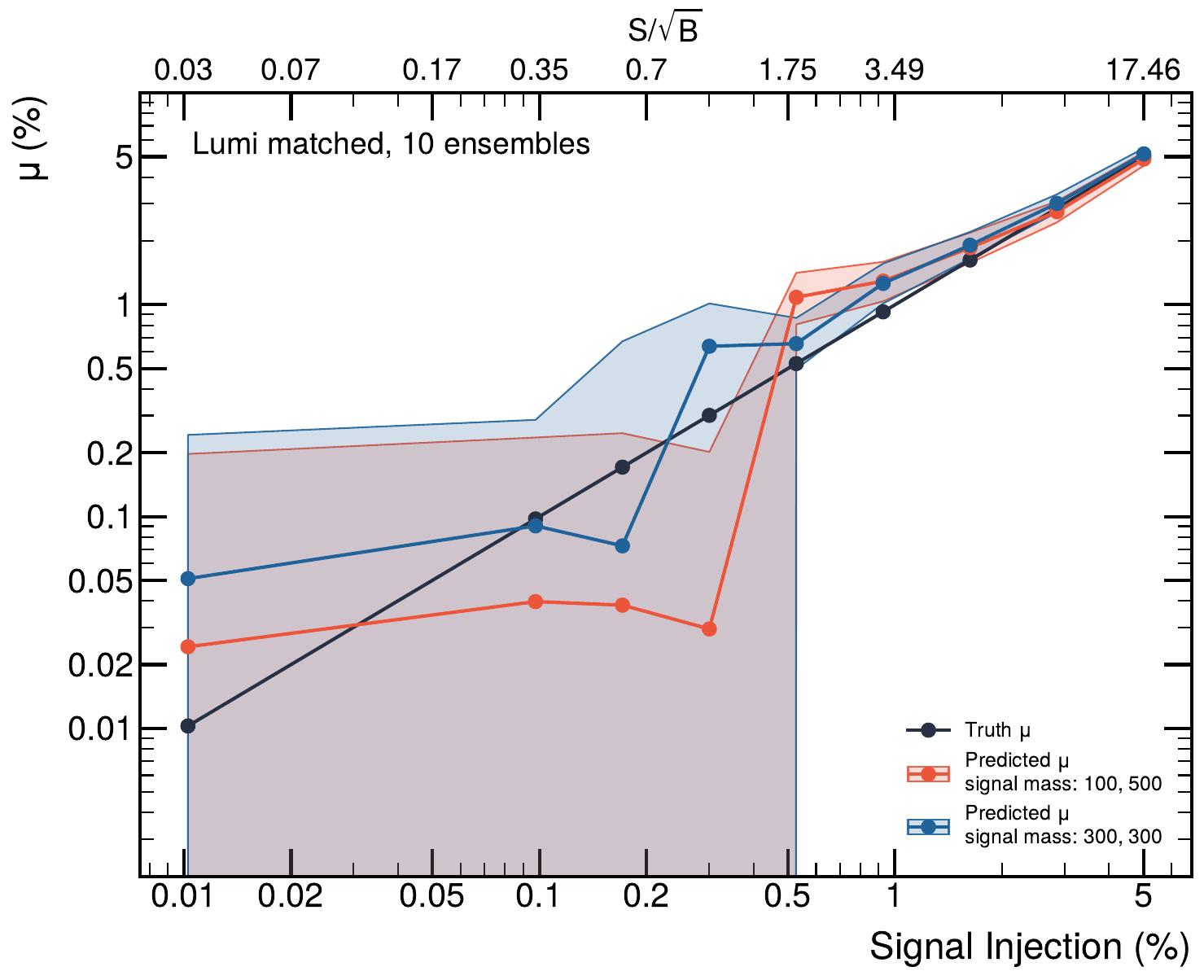}
    \caption{Median estimated signal fraction $\hat{\mu}$ using R-ANODE at signal mass (100, 500) and (300, 300).}
    \label{fig:ranode_comparison_mass}
\end{figure}

\begin{figure}[ht!]
    \centering
    \includegraphics[width=0.95\linewidth]{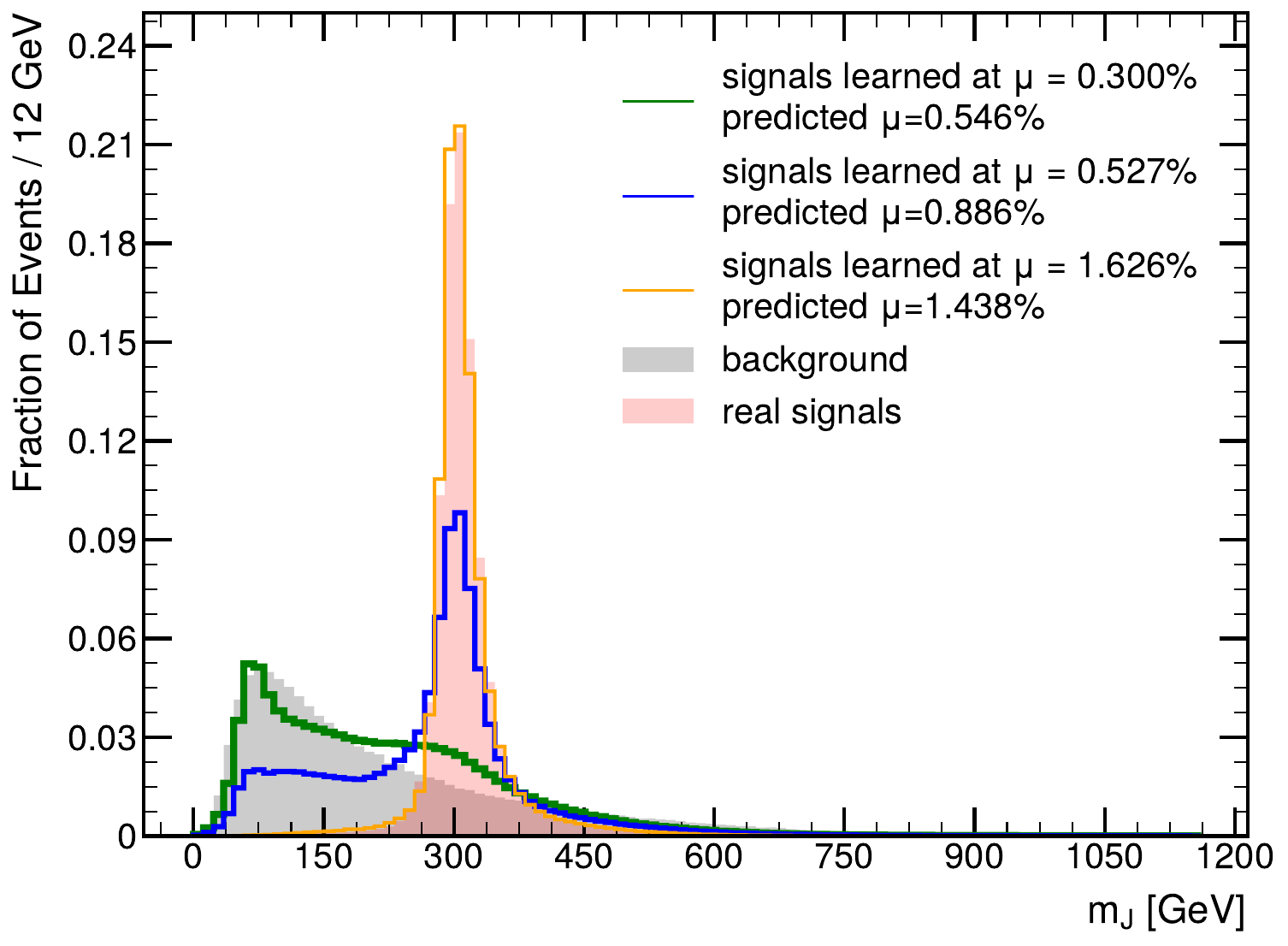}
    \caption{The inferred physical properties of the anomalies for R-ANODE at (300, 300).}
    \label{fig:ranode_mass_contours_300300}
\end{figure}

Since R-ANODE makes no assumption on the signal besides its resonance, the performance is similar across different signal masses. The result at signal mass (300, 300) is compared with previous result at (100, 500) in Fig.~\ref{fig:ranode_comparison_mass}, and the learned physical properties at (300, 300) is shown in Fig.~\ref{fig:ranode_mass_contours_300300}. We find that the performance at signal mass (300, 300) to be slightly better than performance at signal mass (100, 500), which could be due to the signal features at mass (300, 300) being more different than the background.

\end{document}